# Molecular HDD Logic for Encrypted Massive Data Storage

Bingjie Guo[1†], Xinhui Chen[1,2†], An Chen[1†], Jinxin Wang[3†], Wuhong Xue[4†], Tao Wang[4], Zhixin Wu[1], Xiaolong Zhong[1], Jianmin Zeng[1], Jinjin Li[1*], Mao Li[3,6*], Xiaohong Xu[4*], Yu Chen[5*], Gang Liu[1*]

[1]National Key Laboratory of Advanced Micro and Nano Manufacturing Technology, Department of Micro/Nano Electronics, School of Electronic Information and Electrical Engineering, Shanghai Jiao Tong University, Shanghai 200240, China.

[2]Colleague of Information Engineering, Jinhua University of Vocational Technology, Jinhua 321017, China.

[3]State Key Laboratory of Polymer Physics and Chemistry, Changchun Institute of Applied Chemistry, Chinese Academy of Sciences, Changchun 130022, China.

[4]Key Laboratory of Magnetic Molecules and Magnetic Information Materials of Ministry of Education, School of Chemistry and Materials Science, Shanxi Normal University, Taiyuan 030031, China

[5]Key Laboratory for Advanced Materials and Joint International Research Laboratory of Precision Chemistry and Molecular Engineering, Feringa Nobel Prize Scientist Joint Research Center, School of Chemistry and Molecular Engineering, East China University of Science and Technology, Shanghai 200273, China

[6]State Key Laboratory of Supramolecular Structure and Materials, College of Chemistry, Jilin University, Changchun 130012, China.

All correspondence should be addressed to: lijinjin@sjtu.edu.cn, limao@jlu.edu.cn, 3148704620@qq.com, xuxh@sxnu.edu.cn & gang.liu@sjtu.edu.cn.

**ABSTRACT**

Organic memories, with small dimension, fast speed and long retention features, are considered as promising candidates for massive data archiving. In order to satisfy the requirements for ultra-low power and high-security information storage, we design a conceptual molecular hard-disk (HDD) logic scheme that is capable to execute *in-situ* encryption of massive data in pW/bit power-consumption range. Beneficial from the coupled mechanism of counter-balanced redox reaction and local ion drifting, the basic HDD unit consisting of ~ 200 self-assembled Ru$^X$LPH molecules in a monolayer (SAM) configuration undergoes unique conductance modulation with continuous, symmetric and low-power switching characteristics. 96-state memory performance, which allows 6-bit data storage and single-unit one-step XOR operation, is realized in the Ru$^X$LPH SAM sample. Through single-unit XOR manipulation of the pixel information, *in-situ* bitwise encryption of the Mogao Grottoes mural images stored in the molecular HDD is demonstrated.

# INTRODUCTION

The cold storage of massive confidential information in data center coinstantaneously demands high memory density to lower bit cost and high security for anti-hacking purpose. However, the basic units of the traditional hard-disks (HDDs), which are widely used for data archiving, are only capable of encoding binary logic states *via* up- and downward magnetization of the storage media. Operating in a single-bit per cell manner, the data density of such system is severely limited and constrained by the bit size of the magnetic disks. The lack of an effective approach for bit-level encryption of the classified data, on the other hand, deteriorates the security performance of the HDD based storage systems. Organic memories, featured by their small dimension, fast speed and long retention characteristics, are considered as promising candidates of storage-class memory (SCM) for massive data archiving[1-4]. In particular, the versatile conductance states achieved in organic memories not only increase the unit and overall data density of the system, but enable material implemented logic manipulation for high-order information grinding in potential security-concerned applications[5-7].

Since the birth in 2003, gigantic efforts have been devoted to developing various physicochemical mechanisms, including electrochemical redox reaction[8], donor-acceptor charge transfer[9], nanoparticle-based charge trapping[10], ion migration[11], filamentary conduction[12], and conformation reconfiguration[13], to customize the electronic structures and carrier transport dynamics of organic memory materials. Depending on their diverse conductance switching characteristics, both volatile DRAM, SRAM, and nonvolatile Flash and WORM memory behaviors have been demonstrated in organic devices[14-17]. Upon the inclusion of additional metal complexes or multiple redox active choromophores[7,18], coupling between electrochemistry and conformation reconfiguration[19], as well as energy-dependent trapping/detrapping of charge carriers[20], consecutive switching behaviors were achieved in organic memory to increase the unit storage density through multi-bit operation. Governing by the solid-state electronic processes occurred in the organic switching layers, modulation of device conductances also endows the possibility of implementing both linear (e.g., AND, OR, NOT) and nonlinear (e.g., NAND and XOR) logic algorithms for bit-by-bit encryption of the stored information[6,21-25]. Through co-optimizing the compositions, crystalline structures, synthesis approaches and fabrication procedures of organic materials that are compatible with the state-of-the-art CMOS platform over the past decades[26-32], memory devices with cell dimension down to 2 μm and integrated scale reaching 1024 have been realized to prove their theoretical concept for high density storage applications. Nevertheless, these explorations require to switch a bundle of molecules between different material states to exhibit obvious yet distinguishable device resistances. As a compositive result of the organic assemblies, the high device current in the μA to mA range is extremely undesired for data archiving. Developing organic SCM systems with ultralow power characteristics, as well as being capable of executing more expressive logic functors beyond the conventional Boolean operations, will surely intensify their application toward cold storage of massive confidential information in data center.

As a special example of organic digital gadgets, molecular electronics distinguish themselves with extreme potential for ultrahigh density information storage and logic applications[33]. Using a single molecule or a few molecules as constituting components to design and construct functional devices on molecular scale, molecular electronics offer a complementary pathway to tussle the ever-coming Moore's predicament[34-40]. Peculiarly, manipulating electronic characteristics of a trifle of organic molecules may only consume tiny energy, ideally solving the high-power straits of the organic memories[41]. In this contribution, we report the first molecular hard-disk logic scheme that is based on self-assembly monolayer (SAM) of an organometallic complex molecule (OCM). Adopting the conductive-atomic force microscopic (C-AFM) tip with frontend radius of 25 nm as a programming head to write and read the material state encoded digital information, each basic storage unit contains only ~ 200 OCMs of $Ru^XLPH$. Benefiting from the incorporation of redox-active transitional metal cation ($Ru^{x+}$), organic ligands of carbazolyl terpyridine (CTP) and terpyridyl phosphonate (TPP), as well as driftable halogen anions ($Cl^-$) that effectively modulate the energy band diagram and charge carrier transport dynamics of the OCMs, 96 distinct conductance states with linearity approaching 0.98, ultralow power consumption of pW/bit range and symmetric modulation characteristics are demonstrated in the molecular HDD. Combing the associated multi-bit operation and molecular-level spatial resolution potential, data density of the organic storage system can be effectively improved in comparison to traditional magnetic HDDs. More importantly, symmetric switching between consecutive conductance states in the present molecular HDD greatly simplifies the execution of Boolean logic in single storage unit in one step, which is otherwise in-no-way to be realized by asymmetric conductance modulation devices. *In-situ* bitwise encryption of the stored massive data, for instance high-definition replicas of the painted murals in Mogao Grottoes, are showcased through single-unit XOR manipulation with the SAM based molecular HDD.

## RESULTS

**Molecular design and fundamental conductance modulation characteristics**

Being similar to the magnetic hard-disks, molecular HDD uses mechanical programming heads to write digital information into the physicochemical states of the organic functional molecules, and sense the stored data in terms of tiny bit currents (Fig. 1a). Herein, we deliberately design an organometallic complex Ru$^X$LPH, consisting of an organic caping ligand carbazolyl terpyridine (CTP), a redox-active ruthenium cation and an anchoring ligand terpyridyl phosphonate (TPP) to assemble the OCMs onto ITO conductive substrates (Fig 1b and Supplementary Fig. 1-8). Note that electron transfer associated with the coordination bonds between Ru cation and CTP/TTP ligands will lead to partial reduction of the former to a lower oxidized state Ru$^{x+}$ with 2≤x<3. In the meanwhile, the ligands become positively charged as (CTP/TPP)$^{(3-x)+}$. Under external electric field, counter-balanced redox reaction occurs reversibly between the metal cation and organic ligands,

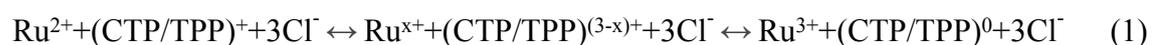

$$Ru^{2+}+(CTP/TPP)^{+}+3Cl^{-} \leftrightarrow Ru^{x+}+(CTP/TPP)^{(3-x)+}+3Cl^{-} \leftrightarrow Ru^{3+}+(CTP/TPP)^{0}+3Cl^{-} \quad (1)$$

whereas the existence of chloride anions inside the OCM always maintain the SAM electrically neutral. More critically, the chloride anions undergo local directional drifting as driven by electric field. The resultant build-in potential, arising from the Cl$^-$ accumulation near either surface of the SAM, modifies the net electric field addressing on the OCMs. As an overall effect of the reversible redox reaction and local anion drifting, the conductances of the OCMs will be modulated continuously. As such, multi-bit, low-power programming of digital information can be achieved through switching the conductance states of OCMs in $Ru^XLPH$ based self-assembly monolayer that serves as storage medium of the molecular HDD.

Fig. 1c and 1d depict the dc current-voltage response and conductance evolution curve of the $Ru^XLPH$ SAM based molecular HDD, which is operated with a conductive-atomic force microscope tip as the programming head to write and read the digital information stored as the molecules' redox and ion accumulation state (Supplementary Fig. 9 and 10). Data writing is executed by applying biased voltage through the programming head onto the OCMs monolayer. As shown in Fig. 1c, the $Ru^XLPH$ OCMs are first scanned in dual-directions, between 0 V to +3.0 V and 0 V to -3.0 V with a ramping step of +0.05 V/-0.05 V. Starting from sampling point #1, the initial conductance of the OCMs is 14.5 nS (Fig. 1d). As the voltage increases, the OCMs conductance drops exponentially to 1.20 nS at 0.1 V and 0.24 nS at 0.5 V (sampling point #2), respectively. Then the I-V curve becomes flat, and the conductance finally reaches 0.06 nS at sampling point #3 with the applied voltage of 3.0 V. Continuous modulation of the OCMs conductance, as expected, may be arising from the oxidization of lower-oxidized state $Ru^{x+}$ ions towards the trivalent form, which in turn modifies the bandgap and conductivity of the organic molecules. Back sweeping to 0 V experiences an immediate steep decreasing of the monitored current, which is accompanied by a polarity reversal of the OCMs' conductance. The conductance is -0.04 nS at sampling #4 of 2.9 V and increases to -0.40 nS at sampling point #5 of 0.5 V. Reversing of the conductance polarity is attributed to the accumulation of $Cl^-$ ions attracted by the positively biased programming head near the $Ru^XLPH$ SAM top surface, wherein the establishment of an additional up-pointing build-in potential can offset the external electric field, therefore influencing the overall effect of the electric field on the charge carrier transport across the organic molecules. At 0 V voltage, the OCM conductance is -36.18 nS. Note that each data point in Fig. 1 and subsequent Fig. 3 represents the average value of 5 individual measurements.

The lineshapes of the I-V and G-V curves recorded during the negative sweeps well resemble that of the positive branch, only minorly differing in the absolute current or conductance values. When swept from 0 V to -3.0 V, the OCMs conductance decreases abruptly from 22.14 nS at 0 V (sampling point #6) to 1.27 nS at -0.1 V and 0.25 nS at -0.5 V (sampling point #7), respectively. Again, the I-V curve becomes flat in the -0.5 V to -3.0 V range, and the conductance reaches -0.06 nS with a polarity reversal at -3.0 V (sampling point #8). Back scanning to 0 V shows a conductance of -0.08 nS at sampling point #9 of -2.9 V and -0.41 nS at sampling point #10 of -0.5 V. Due to the quasi-reversible redox reaction between the metal cation/organic ligands, as well as local drifting of chloride anions, the OCMs' final conductance programmed by C-AFM tip returns to -23.47 nS at 0 V. Note that remarkable conductance windows of $\Delta G = G_{\#2} - G_{\#5} = 640$ pS or $\Delta G' = G_{\#10} - G_{\#7} = 660$ pS are achieved (Fig. 1e), with sampling points #2 and 7 representing the OFF state and sampling points #5 and 10 denoting the ON state. Together with the continuous modulation capability, multi-bit information storage is made possible with the $Ru^X LPH$ samples. Multiplying the monitored current and programming voltage suggests that the peak power consumption is only ~ 690 pW, therefore fulfilling the low-power operation requirements for molecular hard-disks (Fig. 1f). Nevertheless, the spatial resolution of the $Ru^X LPH$ sample is merely limited by the 25 nm frontend radius of the C-AFM tip. As estimated by an atomic force microscope, the number of Ru$^X$LPH molecules constituting a basic molecular HDD unit is ~ 235 (Supplementary Fig. 11). It is therefore reasonable to hypothesize that the extremum bit area and operation power can be shrunk by 235 times to 25 nm/235 = 1.1 Å and 690 pW/235 = 2.94 pW, respectively, depending on promising advances achieved in the technical availability of ultra-miniaturized programming heads for massive storage applications.

**Conductance modulation mechanism**

X-ray photoelectron spectroscopic (XPS) measurements were conducted at room-temperature to confirm the solid-state electrochemistry related conductance modulation mechanism of the Ru$^X$LPH SAM samples. According to the literature[42], the Ru $3d_{5/2}$ binding energies (BEs) of the Ru$^X$LPH molecules are 280.7 eV for Ru$^{2+}$ and 281.3 eV for Ru$^{3+}$ components, respectively. As plotted in Fig. 2a, both the divalent and trivalent Ru ions appear in the OFF state Ru$^X$LPH SAM sample. The co-existence of the Ru$^{2+}$ and Ru$^{3+}$ species are in good agreement with the occurrence of ground state electron transfer through the coordination bonds between the Ru$^{3+}$ cation and CTP/TTP ligands pair, which leads to partial reduction of the trivalent cation to a nominal lower oxidized state Ru$^{x+}$ with 2≤x<3. Integrating along the XPS curve indicates that the Ru$^{2+}$ components share 48.9% of the total metal content, while the Ru$^{3+}$ part occupies 51.1 percent. In addition to the Ru⋯N coordination bond (18.7%) and amine nitrogen species (-N=, 12.2%) at the binding energies of 398.9 eV and 401.4 eV, respectively, the large amounts of N-C components (69.1%) at the BE of 400.9 eV apparently exceed that of the carbazole groups of the CTP ligand (Fig. 2b). It can be ascribed to the newly appeared positively charged nitrogen atoms, as a result of the oxidization of the pyridine nitrogen through electron transfer to the Ru cation upon Ru···N coordination. In accordance with the conductance modulation mechanism that the Ru$^{x+}$ cations become effectively oxidized into Ru$^{3+}$ in the ON state, the relative amount of the latter component increases to 73.2% when the Ru$^X$LPH SAM sample has been stressed with a 10 V voltage. Following the reduction of the initially positively charged terpyridine nitrogen atoms, the N-C content decreases to 40.2%, while the Ru···N coordination bonds and -N= species share 30.3% and 20.4%, respectively. Such redox characteristics of the Ru$^X$LPH SAM sample are in good accordance with its electrochemical properties, which have been already reported in our previous work showing reversible cyclic voltammetric transition with the onset oxidization and reduction potentials of 0.84 V and -1.08 V, respectively[43].

To better understand the redox-related charge carrier transport properties of the Ru$^X$LPH based SAM samples, the static electronic scenarios of both the divalent and trivalent OCMs are investigated through molecular simulation using the Gaussian program package and the density functional theory (DFT)[44,45]. As shown in the left panel of Fig. 2c, the Ru$^{II}$PLH molecule with divalent ruthenium cation shows continuous positive electrostatic potential (ESP) channel in light red color throughout the entire OCM, with the nitrogen atoms of the terpyridine ligands bearing negative ESP spots (blue) and serving as electron accepting centers in the organometallic complex. Due to the cationic nature of the Ru$^{2+}$ ion, it also displays negative ESP potential in the coordination bond. The energy bandgap associated with the lowest unoccupied molecular orbital (LUMO) and highest occupied molecular orbital (HOMO) energy levels, and the dipole moment of the Ru$^{II}$PLH molecule, are 0.626 eV and 10.974 Debye, respectively. Note that electron clouds reside over the entire terpyridine-Ru$^{2+}$-terpyridine coordinated chromophores in the HOMO and LUMO orbitals, which is good agreement with the occurrence of electron transfer between the metal cation and CTP/TTP ligand pairs. When the divalent Ru cation is fully oxidized into the $Ru^{3+}$ form with the terpyridine ligands reduced to the neutral state, obvious reconfiguration of the coordination bonds is visualized (right panel of Fig. 2c). The electrochemical reaction slightly intensifies the HOMO-LUMO energy level difference of the OCM, which increases to 0.635 eV in the $Ru^{III}LPH$ molecule and makes the SAM sample less conductive with relative difficulty of charge carrier transition across the bandgap. $Ru^{III}LPH$ also shows a larger dipole moment of 13.838 Debye, which is favorable for maintaining the as-reached redox and conductance states. As such, modulation of sample conductance exhibits non-volatile nature that is required for long-term data archiving applications. It should also be noted that during molecular simulation the counter anions are fixed to the OCMs through covalent bonds for the ease of calculation convergence. This may lead to minor deviation of the HOMO/LUMO energy levels from the actual values. Nevertheless, the widened bandgap of $Ru^{III}LPH$ will result in device transition into a relatively lower conductance state.

The local drifting of the chloride anions was further assured by the evolution of piezo force microscopic (PFM) phase signals of the self-assembled $Ru^{X}LPH$ monolayer recorded in different redox states (Fig. 2d). After loading external voltages or electric fields onto the SAM sample through the programming head, the polarization features of the monolayer were characterized *in-situ* in PFM mode. As shown in Supplementary Fig. 9, the as-deposited $Ru^{X}LPH$ sample has a nanograined morphology inherited from the ITO substrate. Staying in the pristine state, the mix-valent SAM sample show blue- and yellow-color regions with the respective areal ratios of 47.7% and 52.3%. They correspond to the initial random distribution of chloride anions in the SAM that leads to almost equally aligned up- and downward pointing polarizations. When a voltage of 2.0 V is applied, the color scheme of major area of the Ru$^X$LPH monolayer turns into white, with minor region becomes blue. It suggests that the net-upward pointing polarization of the molecular HDD basic unit is greatly intensified, which is correlated with the upward drifting of chloride anions under the stimuli of external positive electric field. Their accumulation near the surface of the monolayer results in large amounts of negative charges and greatly enhances the attractive interaction between the programming head and the SAM sample. As the programming voltage projected onto the SAM sample increases to 4.0 V, the total area of the upward pointing polarization region extends to 100% with a maximum PFM signal of 0.012, corresponding to a great extent of chloride ion drifting to the head/SAM interface. Further increase in the amplitude of the programming voltages amplifies the PFM signals continuously. In addition, as the spatial separation between the charged Ru$^X$LPH framework and counter anions results in the formation of molecular dipoles, changes in dipole moment of the organometallic complex molecule upon transition between various oxidized states (which is visualized through molecular simulation as discussed above), double assure that the relative positions of the chloride ions with respect to the molecular framework are changed[46-49]. As reported in the literature, local drifting of the counter anion and changes in their relatively position with the molecular framework also influence the width of the molecular energy bandgap[46]. Therefore, the electric field-induced ion drifting, participating in the continuous modulation of the OCMs conductance, is confirmed experimentally in the present

SAM samples.

Based on the above discussion, we try to sketch a phenomenological model that describes the physicochemical process accounting for the multi-bit, low-power conductance modulation of the OCMs in Ru$^X$LPH based self-assembly monolayer (Fig. 2e). At sampling point #1, the oxidized state of the OCM ruthenium cation is between divalent and trivalent, while the chloride anions are distributed evenly across the entire monolayer. The nitrogen atoms on the CTP/TPP ligands become partially oxidized to compensate the positive charges lost by Ru$^{x+}$ upon electron transfer through the formation of Ru$\cdots$N coordination bonds (not shown for ease of illustration). When positive voltage of 0.5 V is applied onto the SAM sample, the lower-valent ruthenium cation becomes oxidized toward the trivalent form with the chloride anion drifting toward the monolayer upper surface. At sampling point #3 with sweeping voltage of 3.0 V, 95% of the Cl$^-$ anions are accumulated near the SAM surface, resulting in a significant upward pointing build-in potential that modify the net electric field addressing on the Ru$^X$LPH based monolayer. As the drifting and accumulation of chloride anions continue during back-scanning, the build-in potential completely offsets the influence of the external electric field at the biased voltage of 2.9 V. Afterwards, the polarity of the net electric field across the monolayer reverses, and the OCMs conductance becomes negative. Further scanning with positive voltage continuously increase the absolute value of the negative conductance of the OCMs, resulting in a large memory window when reading at sampling points #5 and #2 with the same stressing voltages of 0.5 V. Sweeping in the negatively biased branch results in similar conductance modulation and polarity reversing behaviors, as shown at sampling points of #6 to #10, which can be ascribed to the reduction of the $Ru^{3+}$ species to a lower valent form accompanied by the downward drifting of chloride anions towards the $Ru^{X}LPH$ monolayer bottom surface. Note that oxidization of the mixed-valent $Ru^{x+}$ to $Ru^{3+}$ cations only occurs in the positive branch to decrease the conductance of the OCMs, while reduction of the trivalent $Ru^{3+}$ to $Ru^{2+}$ cations only takes place in the negative branch to increase the conductance of the OCMs. Therefore, the symmetric conductance modulation characteristics is mainly attributed to the chloride ion drifting induced build-in potential that continuously modifies the net electric field addressed on the $Ru^{X}LPH$ molecules to deliberately control their charge carrier transport dynamics.

**High-density data storage performance**

A substantial number of conductance states are crucial for enhancing the unit storage density of molecular hard-disks. In order to evaluate its potential for ultrahigh density data storage application, we further assess the current-voltage characteristics of the self-assembled $Ru^{X}LPH$ monolayer in a wider voltage scanning range. As plotted in Fig. 3a, rhombus shape I-V curves with symmetric and continuously expanding hystereses (memory windows) are demonstrated, when the stopping voltage increases from ±0.5 V to ±10.0 V with a ramping step of 0.1 V. For instance, the memory window read at the sampling points #2 and #5 with the stressing voltage of 0.5 V (or at the sampling points #7 and #10 with the stressing voltage of -0.5 V) is 222 pS, when the scan stopping voltages are set as ±0.5 V (Fig. 3b). In case that the scan stopping voltages are 5.5 and 10.0 V, the memory windows are leveraged to 1697 pS and 2907 pS, respectively (Fig. 3c and 3d). Being attributed to the continuous modulations of counter-balanced redox reaction between the $Ru^{x+}$ cations and terpyridine ligands, as well as local drifting and accumulation of the chloride anions, the incremental hystereses are highly favored to increase the number of memory states and thus unit storage density of the Ru$^X$LPH monolayer based molecular HDD. On the other hand, although the application of ±10.0 V voltage to the SAM layer with an estimated thickness of ~ 2.54 nm (length of the organometallic complex molecules read by molecular simulation) will generate a high field of 4×10$^9$ V/m, the possibility that the above discussed conductance modulation characteristics is attributed to electrical breakdown of the organic samples can be safely ruled out. Generally, electrical breakdown is accompanied by thermal pyrolysis related formation of carbon rich conductive filaments in the organic layer. It short-circuits the top and bottom electrodes, which causes large and unswitchable sample currents reaching the compliance level of the measuring instrument. On the contrary, the Ru$^X$LPH based molecular HDD units demonstrate reprogrammable conductance modulation characteristics with pA level sample currents observed during our measurements, confirming that its origin is intrinsic to the changes of organic molecules' properties.

We also calculate the conductance values of the SAM sample at 0.1 V, as illustrated in Fig. 3a and 3e. Upon increasing the magnitude of the voltage stimuli from 0.5 V to 10.0 V, the OCMs conductance increases from 0.4 to 7.3 nS in the positive sweeps. Reversing the scanning polarity leads to similar conductance modulation characteristics between 0.2 and 8.3 nS. Such symmetric conductance tuning takes place in 96 steps with modulation linearity approaching 0.99 and uniformity exceeding 94% (Supplementary Note 4), therefore enabling at least 6-bit storage for high-density data archiving applications. Accordingly, the disk volume required to store the same amount of information with the Ru$^X$LPH monolayer based molecular HDD can be effectively reduced to 16.7% (1/6), in comparison to that of the traditional binary magnetic hard disks. As the 96-state conductance modulation characteristic is demonstrated with the modulating voltages increasing in a ramping step of 0.1 V, further decreasing in the ramping step (e. g. to 0.01 V) may effectively increase the numbers of conductance states in orders of magnitudes approaching that reported in metal oxide based memristor devices[50]. Nevertheless, the 96-state conductance modulation performance is among the best results recorded so far for organic and molecular electronic devices. Moreover, the Ru$^X$LPH SAM exhibits promising stability and reliability of conductance modulation. 10 out of 96 conductance states, which are obtained with the maximum scanning voltages increasing from 1.0 V to 10. 0 V with a ramping step of 1.0 V and read voltage of 0.1 V, are further evaluated for the modulation uniformity and retention capabilities. As plotted in Supplementary Fig. S13, scanning over a single sample for 3 continuous times or scanning over 5 samples reveals that the recorded I-V curves well overlap with each other, showing high cycle-to-cycle and device -to-device uniformities of 98.59% and 96.95% for the OCMs conductance (Fig. 3f and 3g). These conductance states demonstrate good retention performance over 10000 s (Fig. 3h). Although a maximum fluctuation of 15.20% is observed during operation, these 10 conductance states under evaluation are still distinguishable from each other and thus suggest their applicability for data storage usage. Herein, the conductances were read by applying a triangle voltage signal with peak value of 0.1 V and width of 966 ms on the atomic force microscope, wherein the continuous stressing with electrical stimuli may change the sample conductances non-negligibly. In case a short pulse instead of triangle wave is used, a smoother retention curve can be expected reasonably. The conductance modulation performance can be maintained under low temperatures approaching that of the liquid nitrogen (Supplementary Fig. 14), again assuring the stability of the present molecular HDD in various working environments. As mobile ions are frozen at low temperatures, the build-in potential arising from ion migration and accumulation, as well as the net electric field addressed on the SAM layer, can be modified significantly in comparison to that established at room temperature. The degree of sample conductance modulation is attenuated consequently, giving rise to obviously shrunk hystereses in the I-V curves shown in Supplementary Fig. 14.

**Implementation of Boolean and high-order molecular logics**

Encryption *via* bit-wise logic manipulation on the stored information can enhance the security level of massive confidential data. In addition to the feasibility of utilizing conductance modulation to execute logic operators[51-55], its nonvolatile characteristic also allows *in-situ* storage of the computing outputs, which eliminates the necessity of involving additional storage spaces or operations to stock the genuine and newly generated data. As the most general candidates, Boolean logics have been widely demonstrated through on-demand manipulation of conductance in memristive devices[56-58]. Upon one- or two-step input of voltage-based modulating signals, pairing between the initial and final conductance states of a single molecular HDD unit can simply deliver 14 out of 16 Boolean logic operations (Supplementary Fig. 15 and Supplementary Table 1). The remaining XOR operator requires two HDD units to be implemented, while the XNOR gate cannot be achieved technically with the $Ru^XLPH$ molecular HDD (Fig. 4a, 4b and Supplementary Fig. 16). XOR gate is of great importance for information security applications[59-61]. As its two-unit operation methodology based on bistable conductance modulation characteristic of molecular HDD is unable to achieve bit-by-bit encryption of the stored data, more efficient approach should be developed to securely encode the confidential information.

Remembering that beyond the bistable conductance modulation behavior, $Ru^XLPH$ molecules also exhibit bidirectionally symmetric switching characteristics. With such unique feature, we are able to design a single-HDD-unit based one-step algorithm to impart XOR logic operation. As shown in Fig. 4c, programing voltage stimuli of 0 V or 2.5 V are applied through the C-AFM tip and ITO electrode simultaneously as $V_p$ and $V_q$. The difference $V_{p-q}$, which is the actual voltage applied to tune the OCMs conductance, is defined as the logic input. Before operation, the $Ru^XLPH$ molecules reside in the logic "0" state with initial high conductance of 14.5 nS. Then, volage stimuli of $V_p$ and $V_q$ are applied to control the OCMs status. When $V_p$ and $V_q$ equal each other, $V_{p-q}$ and input signal are "0". The OCMs conductance remains unchanged to deliver a logic output of "0". When $V_p$ and $V_q$ are different, $V_{p-q}$ is ±2.5 V and input signal is defined as "1". In case that $V_{p-q}$ is +2.5 V, the low-valent $Ru^{x+}$ species are oxidized into trivalent $Ru^{3+}$ ions. In the meanwhile, the chloride anions drift towards the positively biased programming tip, resulting in an upward pointing build-in potential in the molecular monolayer. As such, the OCMs conductance is reduced to 63.8 pS and output logic state "1". If the $V_{p-q}$ is -2.5 V (in the backward scan), the trivalent $Ru^{3+}$ cations are reduced to the divalent $Ru^{2+}$ form, with the chloride anions accumulated near the SAM/ITO surface to give a downward pointing build-in potential. The OCMs conductance is 92.8 pS and logic output signal is also "1". Fig. 4d and 4e summarize the truth table and simulated results of the single-HDD-unit based XOR operation schemes. In comparison to other conductance modulation related device approaches, the symmetric-switching molecular HDD logic outperforms obviously in terms of unit numbers and operation steps (Supplementary Table 2), not only favoring the reduction of storage space and computing costs but also enabling bit-level encryption of the stored data.

The continuous conductance modulation of the Ru$^X$LPH based molecular HDD also allows the design of high-order logics, which may further simplify the spatial and temporal complexity of computing algorithm[62-64]. For demonstration, we show a computationally complete set of the ternary logic that is closest to the binary counterparts and can be implemented in a single molecular HDD unit, including operators of Plus MAX, Multiply MIN, and Threshold Comparison according to our previous work[65,66]:

$$f(x) = max(p,q) \qquad (2)$$

$$f(x) = min(p,q) \qquad (3)$$

$$f_k(x) = \begin{cases} 2, x = k \\ 0, x \neq k \end{cases} \qquad (4)$$

where *p* and *q* are logic inputs while *k* corresponds to logic states of "0", "1", or "2" (Supplementary Fig. 17 and Supplementary Table 3). Theoretically, all ternary operations can be realized by synthesizing the above complete set *via* logic cascading, which also permits downward compatibility with conventional binary Boolean logics. Beyond the ternary participators, even higher-order operators of quaternary logics can also be demonstrated with a single molecular HDD unit, as shown in Supplementary Fig. 18 and Supplementary Table 4.

***In-situ* encryption of stored Mogao Grottoes Mural picture.**

Finally, we demonstrate the possibility of using molecular HDD as a logic operator protocol for encrypted massive data storage, using the digital image of Mogao Grottoes Mural as an example. A part of the chromatic Bodhisattva mural in Cave 205 is compressed into a 128×128 (16k pixel) image and further decomposed into three monochromatic pictures of the red (R), green (G) and blue(B) primary colors (Supplementary Fig. 19 and Fig. 5a). To make full use of the 96-memory states of the Ru$^X$LPH molecules, the grey-scale values of the monochromatic pixels are divided into 64 levels. Each pixel in these monochromatic images, therefore, can be vividly represented by a 6-bit binary digit. Using traditional binary magnetic HDDs, 18 (6×3) units are required to store the image information of a single pixel, and the entire chromatic mural image consumes 128×128×18=294912 units. Due to the multi-bit memory capacity of the Ru$^X$LPH molecules, each pixel of the molecular HDD only utilizes three units to represent the RGB information. As such, the mural image costs far less of 128×128×3=49152 units to fully store the genuine visual information. In other words, the Ru$^X$LPH based molecular HDD is particularly effective for massive data storage. Supplementary Table 5 and Supplementary Fig. 20 summarize the encoding table and 6-bit pixel greyscale values matrices of the RGB decomposed mural image.

As sketched in the encryption data flowchart of Fig. 5b, the binary pixel intensity information and its monochromatic greyscale value sub-matrices of the mural image are defined as plaintexts. Three sets of encryption key matrices, corresponding to the RGB domains and containing 128×128 numeric strings of 6-bit binary digits each, are generally randomly through the Python function of randint. For each pixel of the monochromatic sub-matrices, bit-by-bit

XOR operation between its plaintext of the 6-bit greyscale value and random key results in a ciphertext of new numeric string. Note that the XOR operation outputs "0" when the binary digits of the input pairs are the same. On the contrary, the operator outputs "1" when the input pairs are different. Reading from Supplementary Files 1-3 and Supplementary Fig. 20-22, the 6-bit greyscale value plaintext, encryption key and ciphertext groups are (110101, 010110, 100011), (000100, 110010, 110110) and (001000, 010000, 011000) for the first pixel of the mural image in the RGB domains, respectively. These encryption operations are simulated by applying voltage stimuli with intensities of 4.0 V (back scan), 5.5 V and 2.9 V to the Ru$^X$LPH based molecular HDD unit to program its conductances to new levels, which are then read and decoded similarly to deliver the ciphertexts (Fig. 5c). Encrypting along the 128×128 monochromatic sub-matrices, following by superimposing the resultant ciphertexts, the mural image can be completely translated into an unreadable mosaic pattern to enhance the data security level (Fig. 5d). The second-round execution of XOR operations between the ciphertext and encryption key matrices then leads to symmetric decryption of the encrypted data. Beneficial from the non-volatile nature of OCMs conductance modulation, both plaintexts and ciphertexts of the mural image can be stored *in-situ* in the same molecular HDD units, greatly conserving the hardware consumption for encrypted massive data storage. Again, it should be emphasized that without the possibility of executing single-unit XOR operation through symmetric conductance modulation, the critical *in-situ* bitwise encryption of the stored data is unable to be realized for the next generation information techniques.

## DISCUSSIONS

In summary, we designed an organometallic complex molecule Ru$^X$LPH consisting of a mixed-valent Ru$^{x+}$ cation and terpyridine based organic ligand pairs. Due to the occurrence of counter-balanced redox reaction between metal cation and ligands, as well as local drifting of chloride anions, the Ru$^X$LPH SAM based molecular HDD exhibits bidirectional, symmetric and continuous conductance modulation with large memory windows. It not only benefits the realization of high-density data storage through multi-bit operation, but allows implementation of *in-situ* bitwise encryption of the stored information, which is highly desired for the storage of massive confidential data in modern society. In addition, control experiments and simulations were conducted to make a double assurance that the observed conductance modulation characteristics are arising from the inherent properties of the organometallic complex molecules. As shown in Supplementary Fig. 23, replacing the transition metal cation Ru$^{x+}$ with Os$^{x+}$ results in rhombus shaped current-voltage curves with similar conductance modulation characteristic, which is accompanied by the variation of molecular energy bandgaps in different oxidized states. As the absolute conductance values of the Os$^X$LPH molecules are thirty times smaller than that of the Ru$^X$LPH molecules, it can be concluded confidentially that the conductance modulation behaviors observed in the molecular HDD are material-specific. The incorporation of other counter anions such as fluoride (F$^-$) and hexafluorophosphate (PF$_6^-$) also influence the changes of molecular orbitals upon transition between different oxidized states (Supplementary Fig. 24), as well as conductance modulation characteristics[46]. Note that since the environmental moisture significantly affects the electrical performance of the Ru$^X$LPH SAM samples (Supplementary Fig. 25), proper encapsulation of the molecular HDD should be carried out for practical applications. In the future, combining the deliberate molecular design cum synthesis strategy, partitioned assembling of customized molecules, and use of flexible substrates, the molecular HDD may even evolve into floppy disks for high-density, high-security portable digital gadgets.

## METHODS

**Synthesis and characterization.** Synthesis details of the organic and organometallic complex molecules are provided in Supplementary Section 1 and Supplementary Fig. 1-8. The $^1$H, $^{13}$C, and $^{31}$P nuclear magnetic resonance (NMR) spectra were recorded using a Bruker AV-500 spectrometer at 25 °C. Matrix-assisted laser desorption/ionization-time-of-flight (MALDI-TOF) and electrospray ionization (ESI) mass spectrometry (MS) were performed using a Bruker Daltonics Autoflex III TOF.

**Electrical measurement.** A nanoscale test structure of Pt/Ru$^X$LPH SAM/ITO was constructed as the basic storage and logic unit of molecular HDD for electrical measurement. The platinum tip of a conductive-atomic force microscope (C-AFM, FastScan Bio) was used as the programming head of the molecular HDD while the ITO substrate served as a universal bottom electrode. All electrical measurements were conducted on the AFM in ambient environment with a relative humidity of ~33% (except for otherwise mentioned). A linear amplifier with single-channel measurement capability and spatial scanning rate of 1.3 Hz is used for the current-voltage measurements in DC voltage sweeping mode. Sample conductance was calculated as the quotient of as-recorded C-AFM current signal divided by programming or reading voltages. The electrical characteristics of the Os$^X$LPH SAM samples were measured similarly.

**Molecular Simulation.** All density functional theory (DFT) calculations were performed using Gaussian 09 package[44,45]. For better comparison between different charged molecules, all states for geometry optimization are in singlet and close shell. The B3LYP functional was used for geometry optimization. The 6-31G(d) basis set was used for the C, H, O, N and P atoms, while LANL2DZ and its corresponding pseudopotential was used for Ru and Os atoms. All geometry optimization was done in the gas phase.

**Logic and encrypted data storage demonstration.** The logic and *in-situ* encrypted data storage operations were investigated using Cadence Virtuoso platform. During simulation, the conductance modulation characteristics of the SAM samples were used as experimental inputs.

## DATA AVAILABILITY

The authors declare that all data supporting the findings of this work are included in the article and Supplementary Information files.

## AUTHOR CONTRIBUTIONS

M. L and G. L conceived the idea. J. W. synthesized and characterized the molecules. B. G., W. X. and P. W. conducted the electrical measurements. A. C. and J. L. performed the molecular simulation. X. C, B. G., Z. W., X. Z. and J. Z. carried out the logic and data encryption simulation. B. G., X. C., A. C. analyzed and visualized the experimental data. B. G., X. C., M. L., X. X., Y. C. and G. L. co-wrote the paper. All the authors discussed the results and commented on the manuscript.

## CONFLICTS OF INTEREST

There are no conflicts to declare.

## ACKNOWLEDGEMENTS

This work is supported by the National Key R&D Program of China (2022YFB4700102). The authors would like to thank Prof. R. S. Stanley and Prof. C. A. Nijhuis for fruitful discussions.

**CAPTIONS FOR FIGURES**

**Figure 1 Design strategy and basic electrical characteristics of Ru$^X$LPH SAM based molecular HDD.** (a) Schematic illustration of the traditional magnetic and conceptual molecular hard-disk. (b) Design strategy of organometallic complex molecule Ru$^X$LPH that may exhibit continuous conductance modulation behaviors. (c) DC current-voltage characteristics, (d) conductance evolution curves, (d) distribution of the ON- and OFF-state conductances and (d) conductance-modulation power characteristics of the Ru$^X$LPH self-assembled monolayer.

**Figure 2 Conductance modulation mechanism of Ru$^X$LPH SAM.** (a) Ru $3d_{5/2}$ and (b) N 1s XPS spectra of the Ru$^X$LPH SAM in OFF and ON states. (c) HOMO, LUMO and ESP distribution of the Ru$^X$LPH molecule with Ru$^{2+}$ and Ru$^{3+}$ cations. The brown, white, blue, green, red and light purple spheres represent carbon, hydrogen, nitrogen, oxygen, phosphor and ruthenium atoms, respectively. (d) Evolution of the PFM phase signals of the Ru$^X$LPH SAM upon being subject to bias voltage of 0 V, 1.0 V, 2.0 V and 3.0 V. (e) A phenomenological model describing the evolution of redox states of the Ru cation as well as the local drifting and accumulation of chloride anions in the Ru$^X$LPH SAM during biased voltage sweepings.

**Figure 3 Multi-level memory performance of Ru$^X$LPH SAM based molecular HDD.** (a-d) DC current-voltage characteristics of the Ru$^X$LPH self-assembled monolayer, recorded with various maximum scanning voltages of ±0.5 V to ±10.0V. (e) 96-state linear modulation of OCMs conductance during the forward and backward scans, recorded with the maximum biased scanning voltages of 0.5 V to 10.0 V, respectively. (f) Cycle-to-cycle uniformity, (g) device-to-device uniformity and (h) retention characteristics of the OCMs conductances.

**Figure 4 Implementations of XOR logic with Ru$^X$LPH SAM based molecular HDD.** (a) Schematic of a XOR logic gate, as well as its implementation with (b) two devices showing traditional redox-related bistable conductance modulation behavior and (c) a single Ru$^X$LPH based molecular HDD unit exhibiting redox and ion drifting induced symmetric conductance switching characteristic. (d) Truth table and (e) simulated results of the as-designed XOR logic operator.

**Figure 5 *In-situ* encryption of a Mogao Grottoes Mural image stored in Ru$^X$LPH SAM based molecular HDD.** (a) Compress and RGB channel generation of a chromatic Bodhisattva mural image in Cave 205 of the Mogao Grottoes. (b) Flowchart of image encryption and decryption through bit-by-bit XOR operations. (c) Simulated data for storing, encrypting and decrypting of the 6-bit greyscale value information of the mural image's first pixel in the RGB channels. (d) Conversion between the genuine chromatic mural image and the encrypted monochromatic and chromatic images.

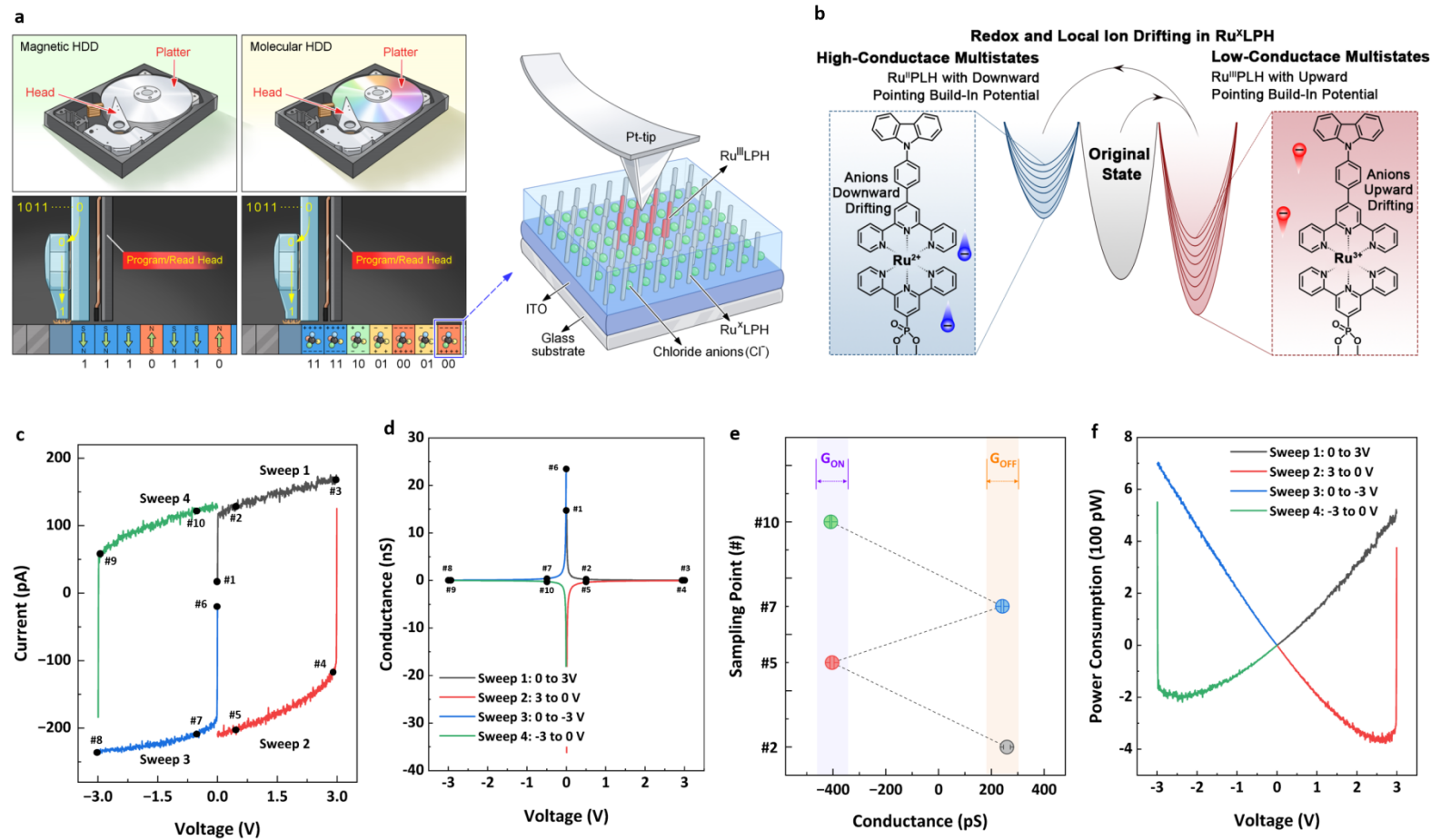

**Figure 1 Design strategy and basic electrical characteristics of Ru$^X$LPH SAM based molecular HDD.** (a) Schematic illustration of the traditional magnetic and conceptual molecular hard-disk. (b) Design strategy of organometallic complex molecule Ru$^X$LPH that may exhibit continuous conductance modulation behaviors. (c) DC current-voltage characteristics, (d) conductance evolution curves, (d) distribution of the ON- and OFF-state conductances and (d) conductance-modulation power characteristics of the Ru$^X$LPH self-assembled monolayer.

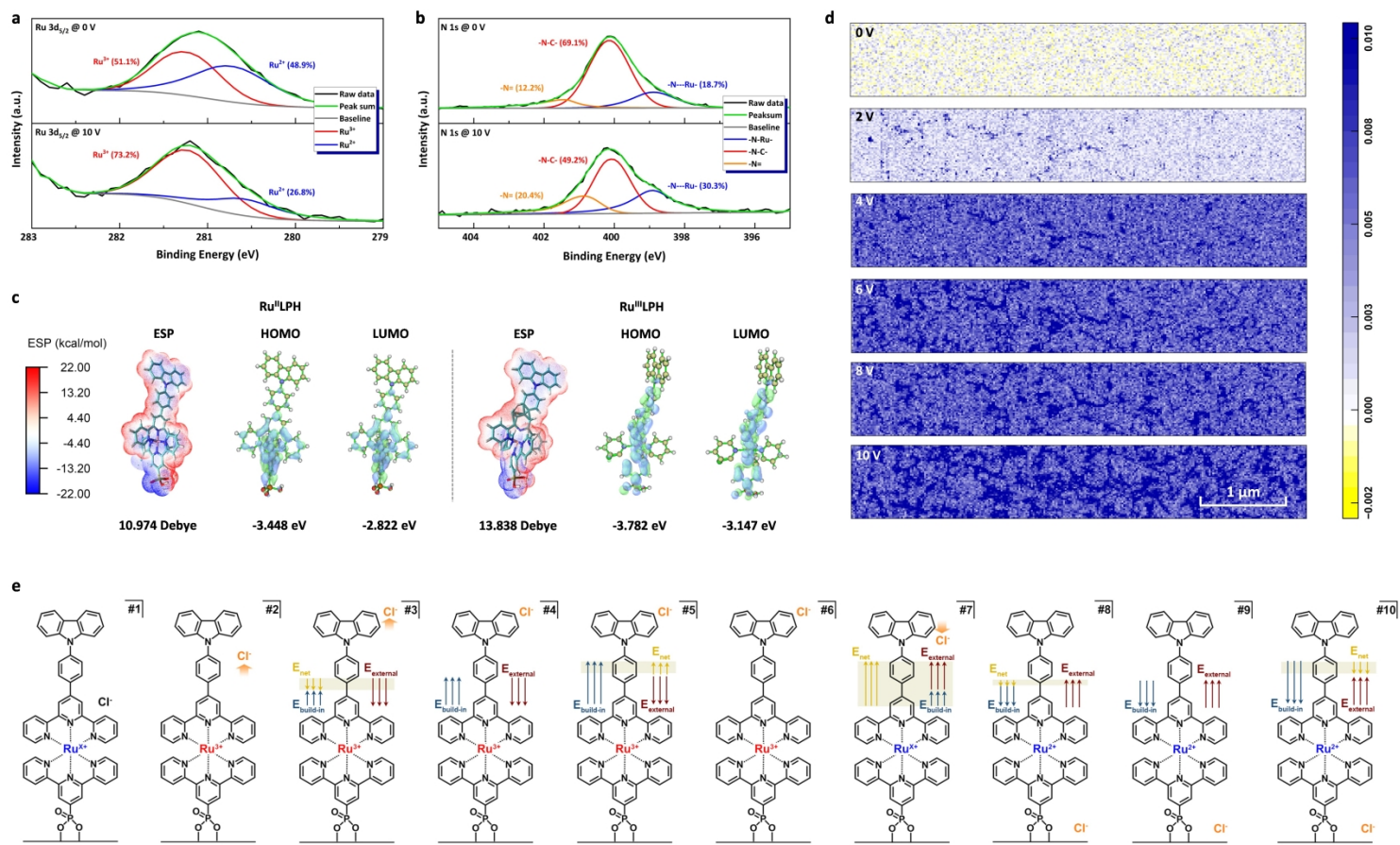

**Figure 2 Conductance modulation mechanism of Ru$^X$LPH SAM.** (a) Ru 3d$_{5/2}$ and (b) N 1s XPS spectra of the Ru$^X$LPH SAM in OFF and ON states. (c) HOMO, LUMO and ESP distribution of the Ru$^X$LPH molecule with Ru$^{2+}$ and Ru$^{3+}$ cations. The brown, white, blue, green, red and light purple spheres represent carbon, hydrogen, nitrogen, oxygen, phosphor and ruthenium atoms, respectively. (d) Evolution of the PFM phase signals of the Ru$^X$LPH SAM upon being subject to bias voltage of 0 V, 1.0 V, 2.0 V and 3.0 V. (e) A phenomenological model describing the evolution of redox states of the Ru cation as well as the local drifting and accumulation of chloride anions in the Ru$^X$LPH SAM during biased voltage sweepings.

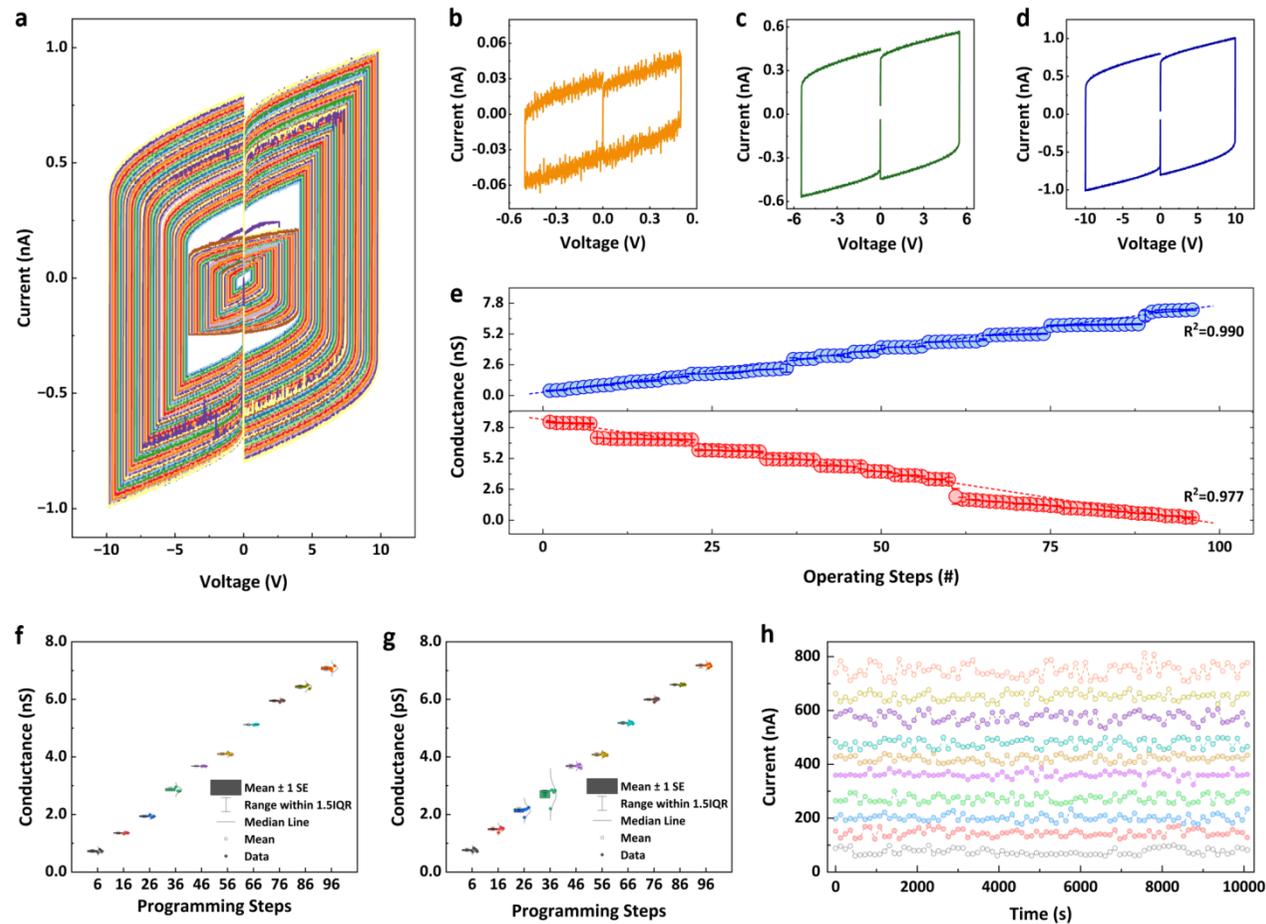

**Figure 3 Multi-level memory performance of Ru$^X$LPH SAM based molecular HDD.** (a-d) DC current-voltage characteristics of the Ru$^X$LPH self-assembled monolayer, recorded with various maximum scanning voltages of ±0.5 V to ±10.0V. (e) 96-state linear modulation of OCMs conductance during the forward and backward scans, recorded with the maximum biased scanning voltages of 0.5 V to 10.0 V, respectively. (f) Cycle-to-cycle uniformity, (g) device-to-device uniformity and (h) retention characteristics of the OCMs conductances.

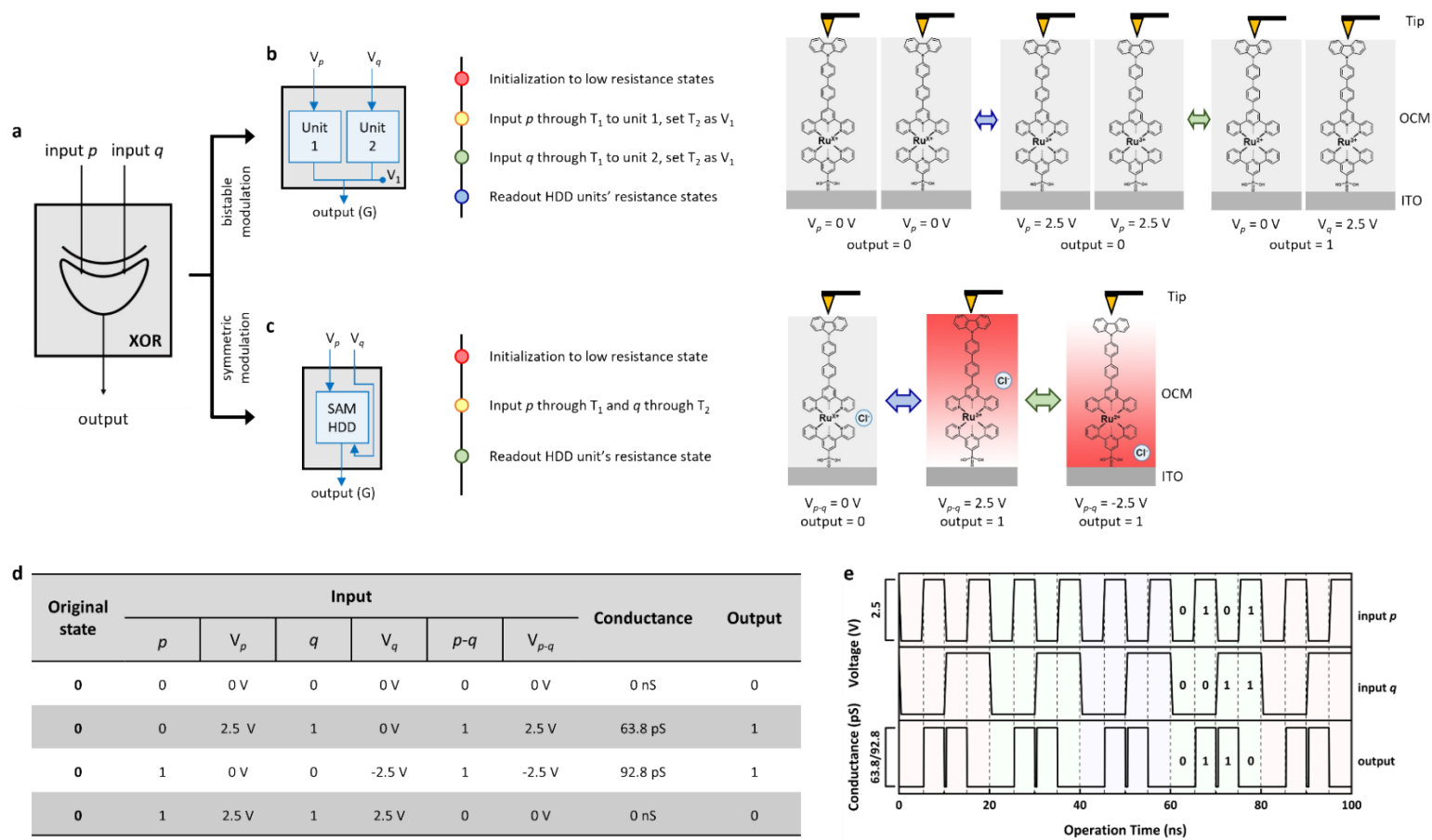

**Figure 4 Implementations of XOR logic with Ru$^X$LPH SAM based molecular HDD.** (a) Schematic of a XOR logic gate, as well as its implementation with (b) two devices showing traditional redox-related bistable conductance modulation behavior and (c) a single Ru$^X$LPH based molecular HDD unit exhibiting redox and ion drifting induced symmetric conductance switching characteristic. (d) Truth table and (e) simulated results of the as-designed XOR logic operator.

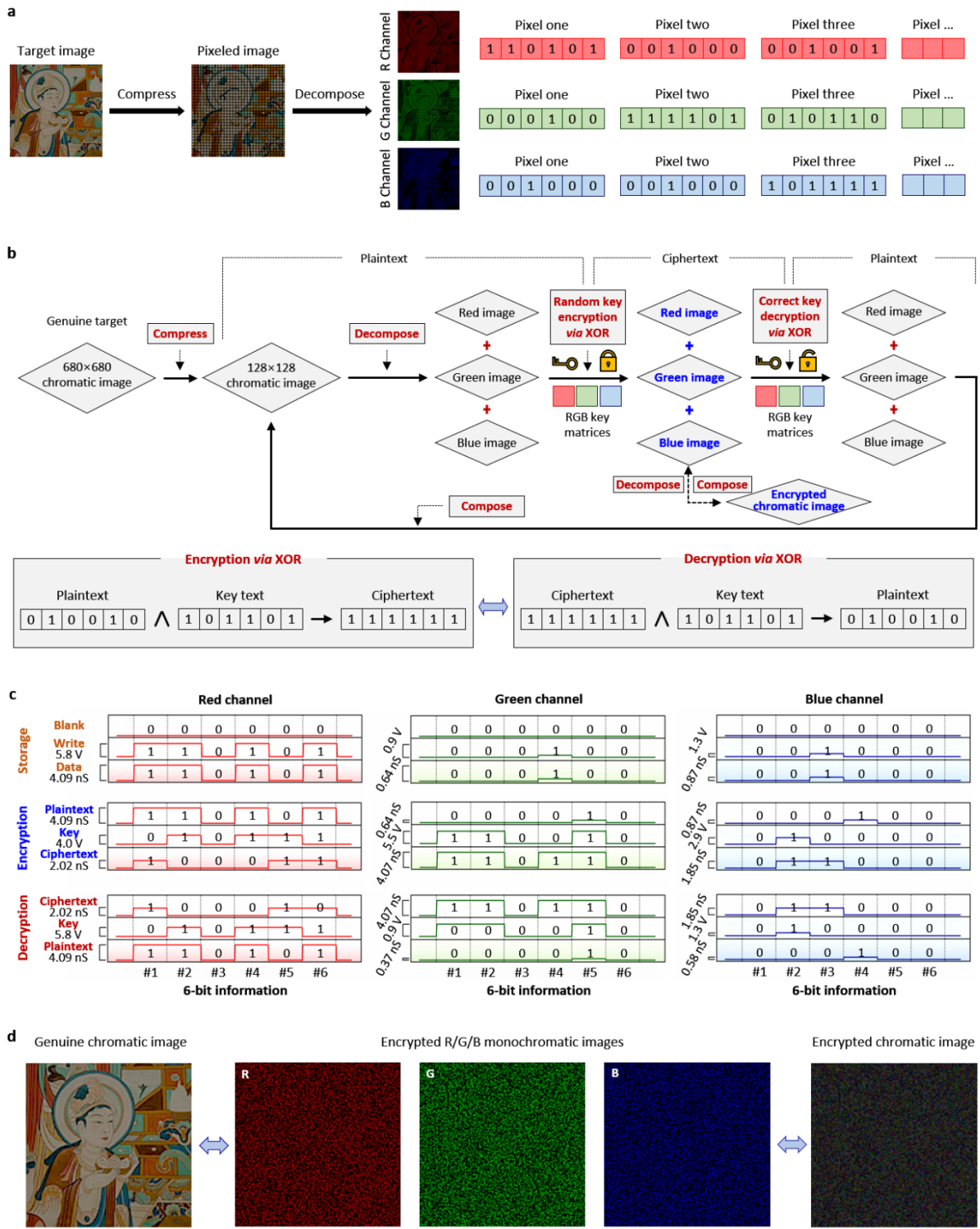

**Figure 5 *In-situ* encryption of a Mogao Grottoes Mural image stored in Ru$^X$LPH SAM based molecular HDD.** (a) Compress and RGB channel generation of a chromatic Bodhisattva mural image in Cave 205 of the Mogao Grottoes. (b) Flowchart of image encryption and decryption through bit-by-bit XOR operations. (c) Simulated data for storing, encrypting and decrypting of the 6-bit greyscale value information of the mural image's first pixel in the RGB channels. (d) Conversion between the genuine chromatic mural image and the encrypted monochromatic and chromatic images.